# *Electronic properties of buried hetero-interfaces of $LaAlO_3$ on $SrTiO_3$*

**Wolter Siemons[1,2,*], Gertjan Koster[1,*], Hideki Yamamoto[1,3], Theodore H. Geballe[1], Dave H.A. Blank[2] and Malcolm R. Beasley[1]**

[1] Geballe Laboratory for Advanced Materials, Stanford University, Stanford, California, 94305, United States of America

[2] Faculty of Science and Technology and MESA+ Institute for Nanotechnology, University of Twente, P.O. Box 217, 7500 AE, Enschede, The Netherlands

[3] NTT Basic Research Laboratories, 3-1 Wakamiya Morinosato, Atsugi-shi, Kanagawa, 243-0198, Japan

[*]Contributed equally to this work

Correspondence should be addressed to g.koster@utwente.nl [G.K.]

**We have made very thin films of $LaAlO_3$ on $TiO_2$ terminated $SrTiO_3$ and have measured the properties of the resulting interface in various ways. Transport measurements show a maximum sheet carrier density of $10^{16}$ $cm^{-2}$ and a mobility around $10^4$ $cm^2$ $V^{-1}$ $s^{-1}$. *In situ* ultraviolet photoelectron spectroscopy (UPS) indicates that for these samples a finite density of states exists at the Fermi level. From the oxygen pressure dependence measured in both transport as well as the UPS, we detail, as reported previously by us, that oxygen vacancies play an important role in the creation of the charge carriers and that these** **vacancies are introduced by the pulsed laser deposition process used to make the hetero-interfaces. Under the conditions studied the effect of $LaAlO_3$ on the carrier density is found to be minimal.**

## Introduction

Recently, Ohtomo and coworkers [1,2] have reported the existence of a conducting electron layer at the hetero-interface between two nominal insulators. This is a remarkable result and has intrigued many researchers in the field [3-6]. It suggests, at a minimum, a very interesting charge transfer system due to valence mismatch of insulators. Specifically, in the work of reference [1,2], the hetero-interfaces were formed from very thin films of $LaAlO_3$ grown on $SrTiO_3$ substrates. $SrTiO_3$ consists of charge neutral SrO and $TiO_2$ layers, whereas the $AlO_2^-$ and $LaO^+$ layers in $LaAlO_3$ are polarized. Thus, when one is deposited on top of the other, a charge layer is expected at the interface, if no other reconstructions take place. Remarkably, Ohtomo and coworkers [1,2] found for a $TiO_2$/LaO interface that this charge layer is conducting and has a sheet carrier density of ~$10^{17}$ electrons/$cm^2$, as inferred from Hall-effect measurements. This carrier density exceeds by two orders of magnitude what one would expect based on simple considerations. The surface polarization expected by simply terminating bulk $SrTiO_3$ with $LaAlO_3$ is half an electron per unit cell, which amounts to $3.2\times10^{14}$ charges/$cm^2$. Even though the original work [1,2] does mention that $10^{17}$ carriers/$cm^2$ is rather unphysical to be explained by an electronic reconstruction, a very natural explanation is found through oxygen vacancies, which was dismissed by those authors. We have already discussed this discrepancy in an earlier paper, where we compared with a variety of physical measurements (near edge x-ray absorption spectroscopy (NEXAS), visible to vacuum UV-spectroscopic ellipsometry (vis-VUV-SE), annealing experiments, transport, and ultraviolet photoelectron spectroscopy (UPS)) samples made under different deposition conditions, and we proposed a model to explain the results [7]. However, in that paper only some of the results of the transport and UPS measurements were presented.

Here we present a fuller account of our results, particularly on the transport and ultraviolet photoelectron spectroscopy (UPS) properties at this interface. In part one, we will show results from resistivity measurements. We show that the magnitude of the sheet density and the mobility of the electrons are a sensitive function of the deposition conditions in ways that suggest that the origin of this large sheet charge density is oxygen vacancies (donating electrons) in the $SrTiO_3$ substrate. Further, we argue that these vacancies are introduced by the pulsed laser deposition (PLD) process used by us and all previous authors to grow these hetero-structures. We will close this part by discussing localization effects at low temperatures and hysteretic behavior in the resistivity measurements. In part two we will discuss the UPS spectra of the buried interfaces with the highest sheet carrier densities, i.e. the films made at the lowest pressures. We will look at the number of states that appear at the Fermi level and at their dependence on deposition conditions, where several interesting trends are observed. Before we discuss the results we turn to the experimental parameters.

## Experiment

All of the films reported here were grown using PLD in a stainless steel vacuum chamber with a background pressure of $10^{-9}$ Torr. A KrF excimer laser produces a 248 nm wavelength beam with typical pulse lengths of 20-30 ns. A rectangular mask shapes the beam, and a variable attenuator permits variation of the pulse energy. The energy density on the target has been kept at approximately 1.2 $J/cm^2$ unless specified otherwise. Before each run, the rotating $LaAlO_3$ target was pre-ablated for two minutes at 4 Hertz. The $LaAlO_3$ layers were grown on $TiO_2$-terminated $SrTiO_3$ substrates at 815 °C. The HF chemical treatment needed to form $TiO_2$-terminated $SrTiO_3$ substrates is described elsewhere [8].

Various parameters were varied from run to run. These included laser repetition rate, number of laser pulses, mask size and post-deposition treatment. The effects of these variations are important and will be discussed below. In addition, to obtain consistent results, we found it necessary to use a standard post-deposition treatment for samples grown at low oxygen pressures ($10^{-6}$ Torr) in which both the oxygen flow and the substrate temperature were quenched immediately. This treatment is believed to prevent the very thin $LaAlO_3$ layer from diffusing oxygen during cooling; this way it is possible to look at the interface as grown. At higher oxygen pressures ($10^{-5}$ Torr), such treatment was not necessary and samples were cooled down in deposition pressure. During growth, RHEED was used both to determine the amount of material deposited (by counting the periods of intensity oscillations in the standard way) and to monitor the morphology of the sample. RHEED intensity oscillations reveal that typically 120 laser pulses are required to grow a monolayer of $LaAlO_3$. After deposition, the samples were moved in situ into an adjacent photoemission analysis chamber ($<5\times10^{-10}$ Torr base pressure). Electrical transport properties were measured ex situ with a Quantum Design Physical Property Measurement System using the Van der Pauw geometry, taking appropriate precautions to avoid photo-induced carriers.

## Part one: Transport

Here, we discuss transport properties of two classes of samples. The first was prepared under relatively low oxidation conditions ($10^{-6}$ Torr, as measured with a hot cathode ion gauge), resulting in a high number of carriers. The second was deposited and cooled at higher oxidation conditions ($2\times10^{-5}$ Torr), resulting in a reduced carrier density. In Figure 1 we show the temperature dependence of the resistivity for representative samples of the two classes. Together with accompanying Hall measurements, we deduce a sheet carrier density of $\sim10^{16}$ $cm^{-2}$ and a mobility of $10^4$ $cm^2$ $V^{-1}$ $s^{-1}$ (at 4 K) for the samples that are made at low pressure and a sheet carrier density of $\sim10^{13}$ $cm^{-2}$ and a mobility of $10^3$ $cm^2$ $V^{-1}$ $s^{-1}$ (at 4 K) for the samples made at higher pressure. These findings are very much in line with what has been reported by others and demonstrates directly that the oxygen pressure during deposition clearly affects the transport properties of these hetero-structures.

Now we focus on the transport data for films made at low oxygen pressure ($10^{-6}$ Torr), since these display the most extraordinary densities and mobilities. Note that these samples are also discussed in the next section on UPS measurements. The sheet resistance and Hall effect of the films were measured as a function of temperature. The results for the 1, 3 and 5 monolayer (ML) samples are shown in Figure 2. Note that the 1 ML thick sample was additionally oxidized after deposition whereas the other samples were not. In any event, the sheet resistance (see inset of Figure 2a) of all samples is strikingly low. Also, both the temperature dependence and the absolute values of the sheet resistance are in good agreement with those reported by Ohtomo *et al*. [[1,2]]. The Hall-effect data shown in the inset of Figure 2b are also similar to those of reference [[1,2]]. The electron mobilities and sheet carrier densities derived from these data are shown in Figures 2a and b. The inferred sheet carrier density is $1.5 \times 10^{16}$ $cm^{-2}$ for the 1 ML thick film (post annealed in oxygen), and $7.6 \times 10^{15}$ $cm^{-2}$ for the 3 and 5 ML samples at 4 K.

Next, we turn to the interfaces made under high oxidation conditions. From resistivity and Hall measurements we arrive at a sheet carrier density of $\sim10^{13}$ $cm^{-2}$ and a mobility of $10^3$ $cm^2$ $V^{-1}$ $s^{-1}$ (at 4 K), as expected lower than the previous case, this is shown also in Figure 1. In addition, an interesting temperature hysteresis effect is measured for the films with low carrier densities, this is shown in Figure 3. The amplitude of these features increases for thinner $LaAlO_3$ overlayers: so far a maximum effect has been observed for 4 unit cell layers for as deposited films (fewer than 4 layers deposited resulted in insulating samples for these deposition conditions, which has also been observed by Thiel *et al.* [4]); and they also increase with magnetic field. The temperature at which they occur is always the same though. These features appear not to be stable over time: when during warm up the temperature is fixed to the maximum of one of the features, the resistance of the sample slowly drops over time until it reached the resistance the sample had during cool down. The time over which this happens is in the order of hours.

## Part two: Photoemission

Of the samples made under low oxygen pressure, of which the transport properties were discussed in the previous section, UPS spectra were taken. The angle-integrated UPS data for hetero-structures with different $LaAlO_3$ layer thicknesses are shown in Figure 4. These UPS spectra were taken for hetero-structures with $LaAlO_3$ films ranging in thickness from a half to three unit cell layers, as determined by RHEED. Note that as the thickness of $LaAlO_3$ is increased, the relative intensities of the oxygen 2*p* double-peak structure (at about 5 eV and 7 eV, respectively) change in a systematic fashion. Unfortunately, excessive charging has prevented us from taking a useful ‘0 ML case’ spectrum, i.e., undoped $SrTiO_3$ (instead we have added a spectrum of a Nb doped $SrTiO_3$ substrate). On the other hand, this charging confirms directly that the $SrTiO_3$ is initially insulating. In addition, by extrapolation, our data seems compatible with known fully oxidized $SrTiO_3$ UPS data (see for example Henrich and Cox’s book [[9]]).

To check the conductivity of the sample one has to measure carefully around the Fermi level to check for states being present. As an example, an expanded version of the UPS data near the Fermi energy is presented in Figure 5 for a representative sample, both as deposited and after additional oxidation. This particular sample had a 1 ML thick $LaAlO_3$ layer and was deposited at a repetition rate of 1 Hz in an atmosphere of $10^{-6}$ Torr molecular oxygen. The additional oxidation was carried out by exposing the sample to 6000L of oxygen at 150 °C. The oxidation of the sample had no discernable effect on the UPS spectrum outside the region shown in Figure 5.

Most importantly, the UPS data show the existence of electron states extending up to the Fermi energy, indicating the existence of a conducting charge layer at the interface (the conducting nature of this interface was confirmed directly by transport measurements presented below). Moreover, the size of this density of states is sensitive to the oxidation history of the sample, being lowered by further oxidation. At a more quantitative level, the photoemission signal at the Fermi level for the 1 ML sample is roughly 1000 times smaller than the O 2p peak. This ratio is comparable to that measured in photoemission for $Sr_{0.95}La_{0.05}TiO_3$, confirming a high carrier density at the surface [[10]]. Earlier work by Takizawa *et al.* also revealed high densities of states observed by photoemission for related samples of $LaTiO_3/SrTiO_3$ hetero structures prepared at low oxidation conditions [11].

In order to compare the resulting density of states produced using various deposition conditions, we subtracted from the UPS data a background function given by the overall trend in the data from -3 eV to +2 eV (see the green dotted line in Figure 5 for the sample shown). An example of this difference spectrum is shown as the inset in Figure 5 for the case of one monolayer deposited at 1 Hz in $10^{-6}$ Torr of oxygen. Finally, in order to better statistically average the data, we integrated the area under the difference spectra. Physically, this integrated density of states corresponds to the number of carriers per unit volume in the layer. These values for various samples are compared in Figure 6.

In Figure 6a, we see that the integrated density of states decreases monotonically with the thickness of the $LaAlO_3$ for otherwise identical samples. This is consistent with the states existing at the interface and the usual escape depth arguments, i.e., the fall off is roughly exponential with a characteristic mean free path of ~8 Å. Samples with thicknesses greater than 5 ML yielded no appreciable photoelectron signal, and for an order of magnitude greater thicknesses, insulating behavior and its associated charging effects prevented UPS study.

Figure 6b shows the effect of varying the PLD laser repetition rate. The data suggest an optimal repetition rate of ~1 Hz for the number of interface states generated. Repetition rates over 5 Hz did not show any discernable UPS signal.

Finally we discuss some special cases. When exposing the 1 ML film deposited at 1 Hz to $10^{-5}$ Torr oxygen for ten minutes at 150 °C the spectral weight at the Fermi energy is reduced, see Figure 5 and Figure 6a (red open triangle). Signal reduction is also observed when the pressure quench is not applied and thus the sample is cooled in $10^{-6}$ Torr of oxygen (Figure 6b, open circle) or similarly when the laser energy per pulse is lowered from 1.2 to 0.8 J $cm^{-2}$ (Figure 6b, solid triangle). Samples grown at higher oxygen pressure ($10^{-5}$ Torr) show no discernable signal at the Fermi level.

## Discussion

Summarizing the results in part two: as we argued in [7], the experimental results presented above strongly suggest that the origin of this high charge density is associated with oxygen vacancies in the $SrTiO_3$, despite the fact that prior to deposition such vacancies are not present. The existence of states at the Fermi level is shown explicitly by the UPS data, as is the fact that the number of states can be reduced by oxidation. The various dependencies of the results on deposition conditions are consistent with this view. The data demonstrate clearly that the number of states is sensitive to deposition conditions. A natural interpretation of these data is that competing processes are present. For example, deposited material at the surface may need some time to crystallize, which we estimate to be on the order of one tenth of a second [12], and for the lowest repetition rates re-oxidation occurs. Another is the sputtering of oxygen from the surface being balanced by growth of the $LaAlO_3$ over-layer and anion vacancies formed deeper in the $SrTiO_3$ [13] through oxygen diffusion. Further evidence for oxygen vacancies and a discussion on suppressing them was presented by ourselves in a recent paper [7] as concluded from NEXAS and vis-VUV-SE measurements and others [3] based on cathode-luminescence. Similarities in transport between the $LaAlO_3/SrTiO_3$ interface and low energy Ar sputtering were found by Reagor and Butko [14] and suggest that particles generated by the PLD process can indeed cause the $SrTiO_3$ to become conducting.

The effect of oxygen vacancies is also clearly reflected in the transport measurements, where sample properties depend on deposition conditions. For samples made at low oxygen pressures we showed remarkably high sheet densities. The inferred mobilities (~$10^4$ $cm^2$ $V^{-1}$ $s^{-1}$) at 4 K are also high, suggesting that the charge donors of the interface carriers are well removed from the interface region itself, as in modulation doping in semiconductor hetero-structures, which was described elsewhere [[7]]. For samples made at higher oxygen pressure we find temperature hysteresis. Extra features were shown to appear at fixed temperatures, which tend to relax slowly over time. These increases in resistance occur at the same temperatures at which increases in the dielectric constant are observed [[15],[16]]. Although they have observed these dielectric relaxations in La doped $SrTiO_3$, Yu and coworkers [[15]] clearly show that they are related to oxygen vacancies in the sample. They anneal in oxygen and/or air at high temperatures (1100 °C) and find that the relaxations disappear. We cannot anneal at such high temperatures since there is a high possibility of damaging the interface. They show as well that at higher La doping values these oscillations disappear. Along the lines of the model we presented in an earlier paper [[7]] we can argue that the location of the electrons is very dependent on the dielectric constant of the material. The effect of the dielectric relaxations will therefore be more pronounced when carrier densities in the undamaged $SrTiO_3$ are low compared to the reduced region. Similar relaxation times have been observed by Thiel *et al.*, in a field effect structure based on these hetero structures [[4]].

In that same paper a critical thickness was observed for a sample to be conductive: samples with a $LaAlO_3$ layer thickness of fewer than 4 ML were found to be insulating. Interestingly, the hysteresis also appears to be maximal for samples with an over-layer thickness of 4 ML. What causes this critical thickness has not been determined yet. If one assumes an intrinsically doped interface though (due to the polarized nature of the $LaAlO_3$ [1,2]) the potential that is created can put electrons in the conduction band. For this effect to occur it needs to be energetically favorable to put an electron in the conduction band and for that to happen the potential over the $LaAlO_3$ layer needs to be larger than its bandgap. We calculated that the jump in potential per unit cell is ~3.8 eV, from $\Delta V=8\pi e^2/\varepsilon a$, which is quite large compared to the bandgap of 5.6 eV. This would mean that after two unit cells there is enough of a potential to put electrons in the conduction band. For real films more layers are probably needed due to the presence of defects. Also note that if some oxygen vacancies are present in the $SrTiO_3$ the diverging potential in the $LaAlO_3$ is attenuated due to the charge already being present in the $SrTiO_3$. For the samples grown at low pressures the high carrier density is present even with only 1 ML of $LaAlO_3$, effectively ruling the charge transfer mechanism out as the source of it.

Even though we see signs of oxygen vacancies in all our films their density can be strongly reduced by choosing the correct deposition conditions. This leads us to suggest that for sufficiently low oxygen vacancy densities two mechanisms could be at play creating charge carriers (oxygen vacancies and the charge transfer due to the polarize nature of the $LaAlO_3$), which seem to balance one another and their effects add up in the regime where both mechanisms of the same order of magnitude. From our data we cannot tell at which electron density this is the case.

Finally, in order to clarify the observed high mobilities in the presence of a large number of defect oxygen vacancies, we showed that a simple model predicts that most of the charge carriers in fact are moved away from the defect layer where they originated and therefore experience less scattering, published elsewhere [7].

## Acknowledgements

One of us (G.K.) thanks the Netherlands Organization for Scientific Research (NWO, VENI). W.S. thanks the Nanotechnology network in the Netherlands, NanoNed. We also wish to acknowledge helpful discussions with T. Claeson, R.H. Hammond, A.J.H.M. Rijnders and G. Lukovsky. This work was supported by the DoE BES with additional support from EPRI.

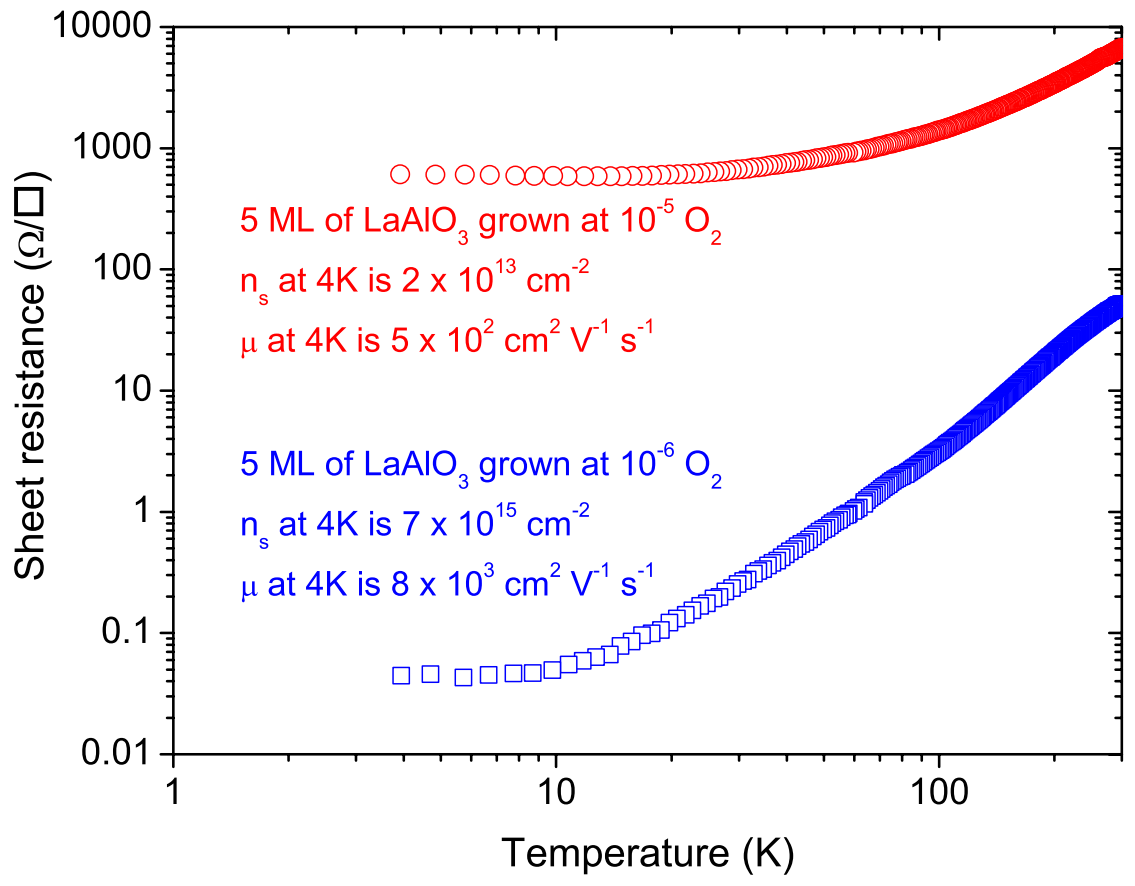

Figure 1. (color online) Resistance as a function of temperature for two 5 ML thick $LaAlO_3$ on $SrTiO_3$ samples, deposited at different oxygen background pressures: one sample deposited at $10^{-5}$ Torr $O_2$ (red open circles) and one at $10^{-6}$ Torr $O_2$ (blue open squares). Also given are some important parameters of both films at 4 K.

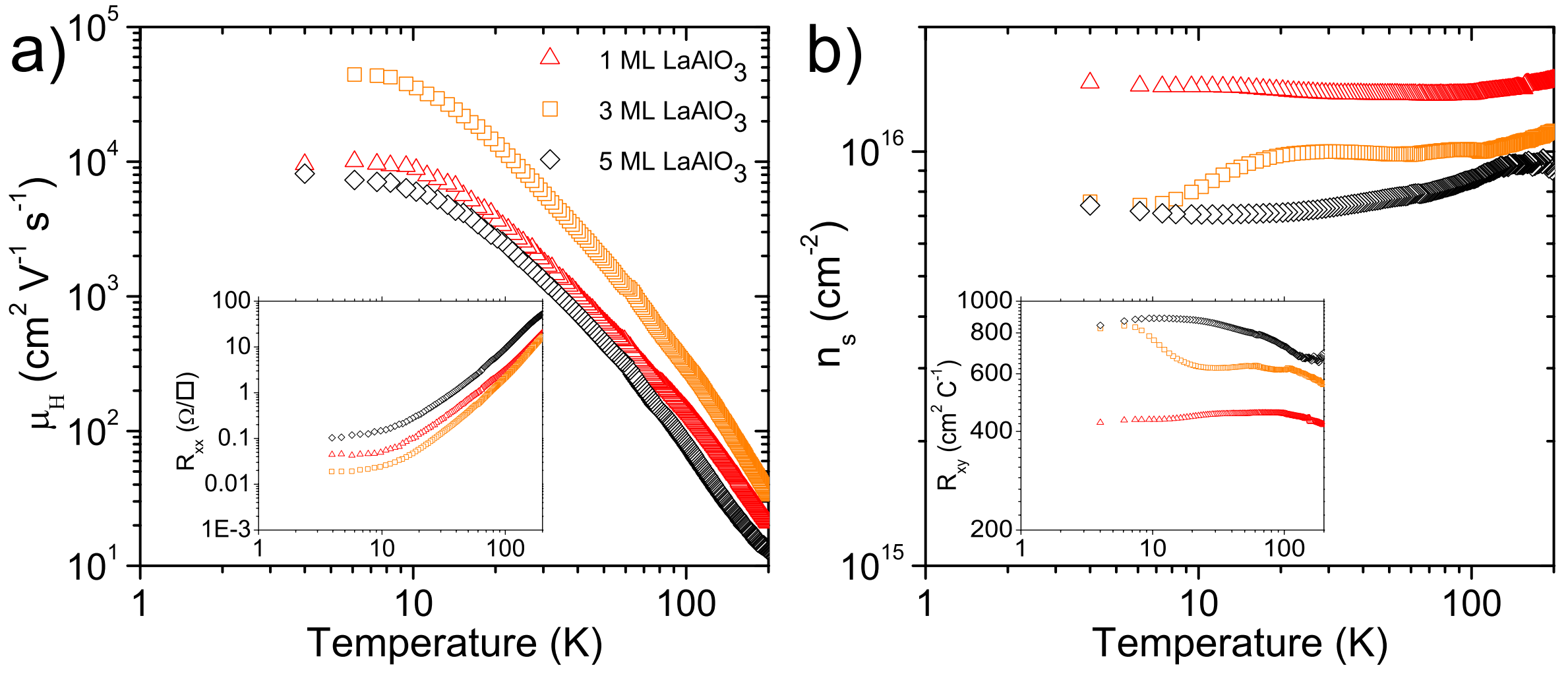

Figure 2. (color online) Transport measurements and analysis of films grown at low oxygen pressure ($10^{-6}$ Torr). a) The mobility of the charge carriers, mH, as a function of temperature of the one (red triangles), three (orange squares) and five (black diamonds) monolayer thick films of $LaAlO_3$ on $SrTiO_3$; the one monolayer sample has been oxidized (6000 L). The sheet resistance, $R_{xx}$, is shown in the inset. b) The sheet carrier density for the same samples calculated from the Hall resistance data ($R_{xy}$, see inset)

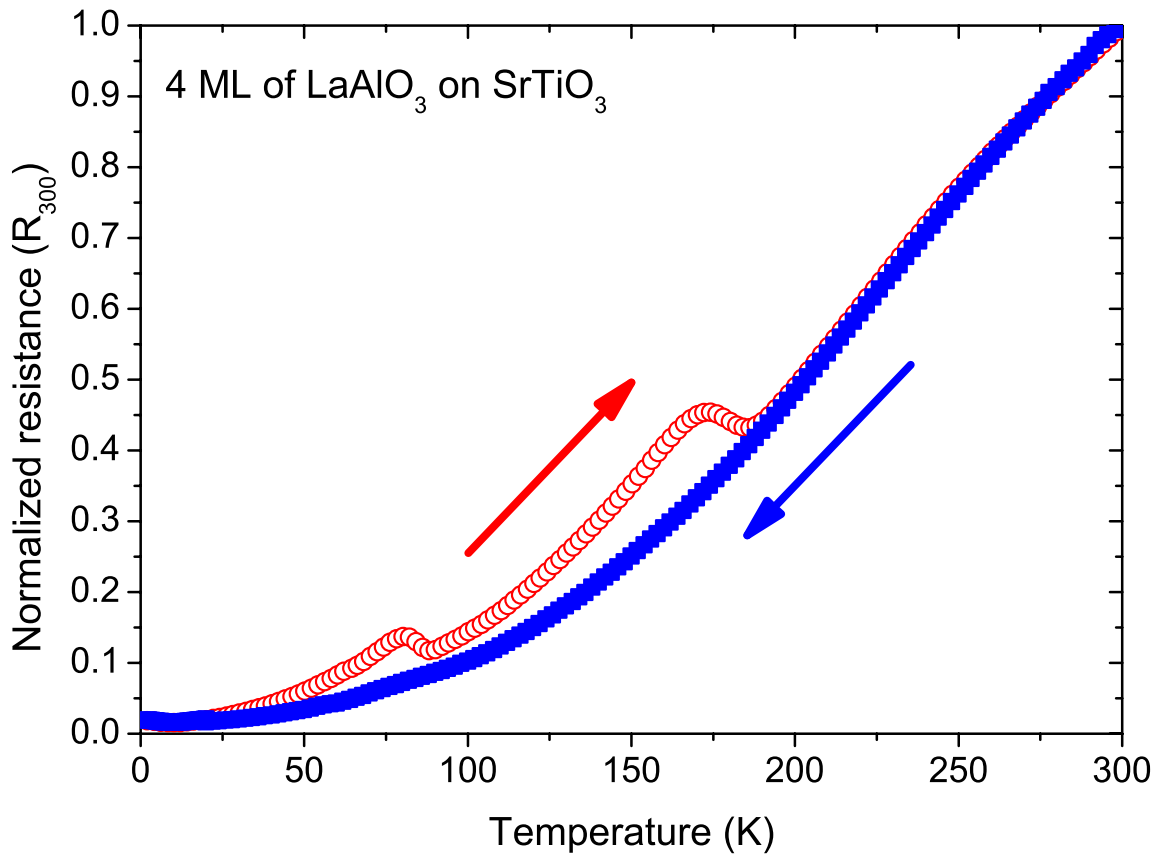

Figure 3. (color online) Extra features show up in the resistance as a function of temperature plots during warmup (red open circles) than during cooldown (blue solid squares). These increases in resistivity have been observed in multiple samples and always occur at the same temperatures.

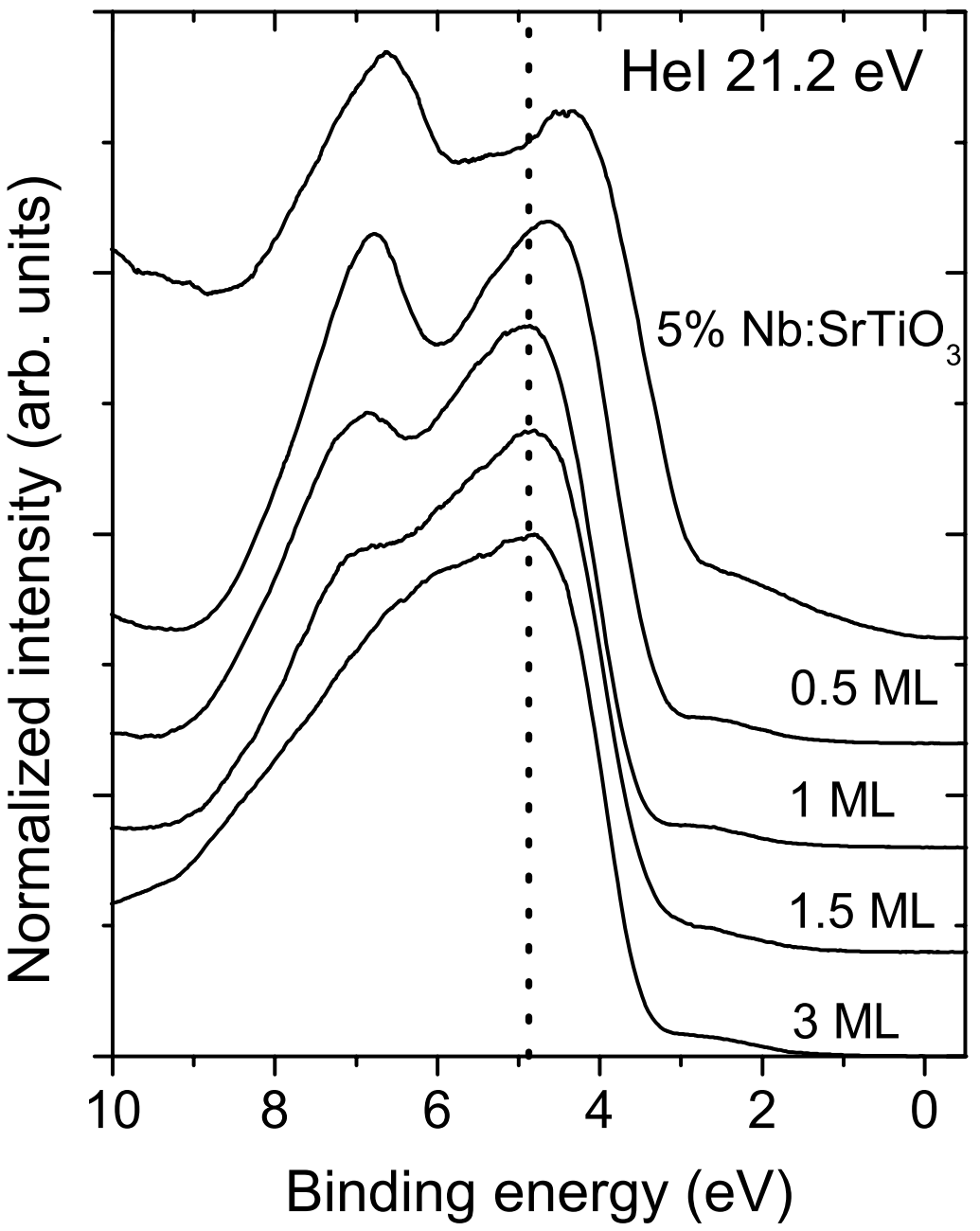

Figure 4. Normalized UPS (HeI 21.2 eV) wide scans for samples with different $LaAlO_3$ thicknesses on $SrTiO_3$. A clear trend is visible as the spectrum changes from bare $SrTiO_3$ (5% Nb doped $SrTiO_3$ was used to avoid charging of the sample) to 3 ML of $LaAlO_3$. The thickness of the $LaAlO_3$ films was varied from 0.5 to 3 unit cell (UC) layers based on RHEED oscillations. The dotted line is a guide to the eye. These spectra were given an arbitrary offset for clarity.

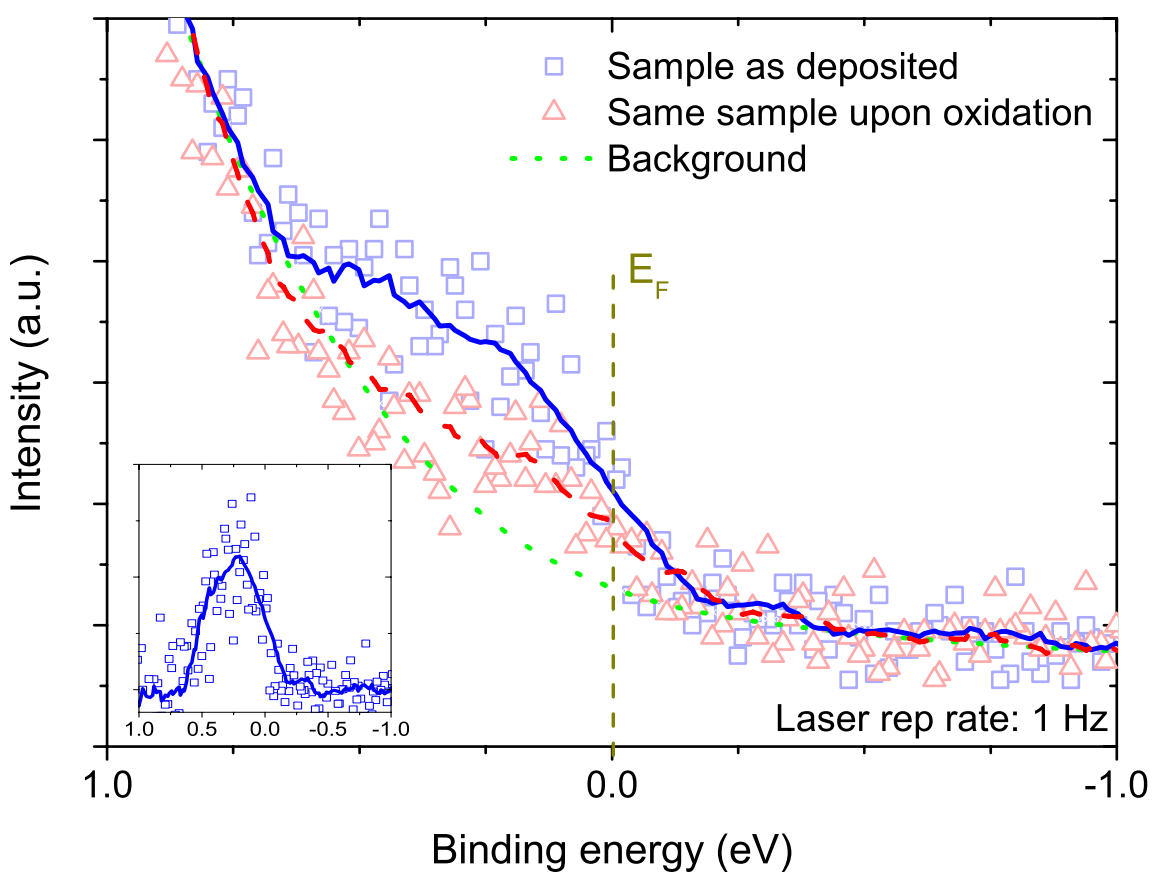

Figure 5. (color online) UPS near the Fermi level. Raw UPS HeI spectra of one monolayer thick $LaAlO_3$ layer on $TiO_2$ terminated $SrTiO_3$ prepared at 1 Hz (blue open squares, fitted with a solid blue line) and the same sample after exposure to ~6000L of oxygen (red open triangles, fitted with a dashed red line). Inset: 1 ML sample before oxidation upon subtraction of background indicated by the dotted green line in the main figure.

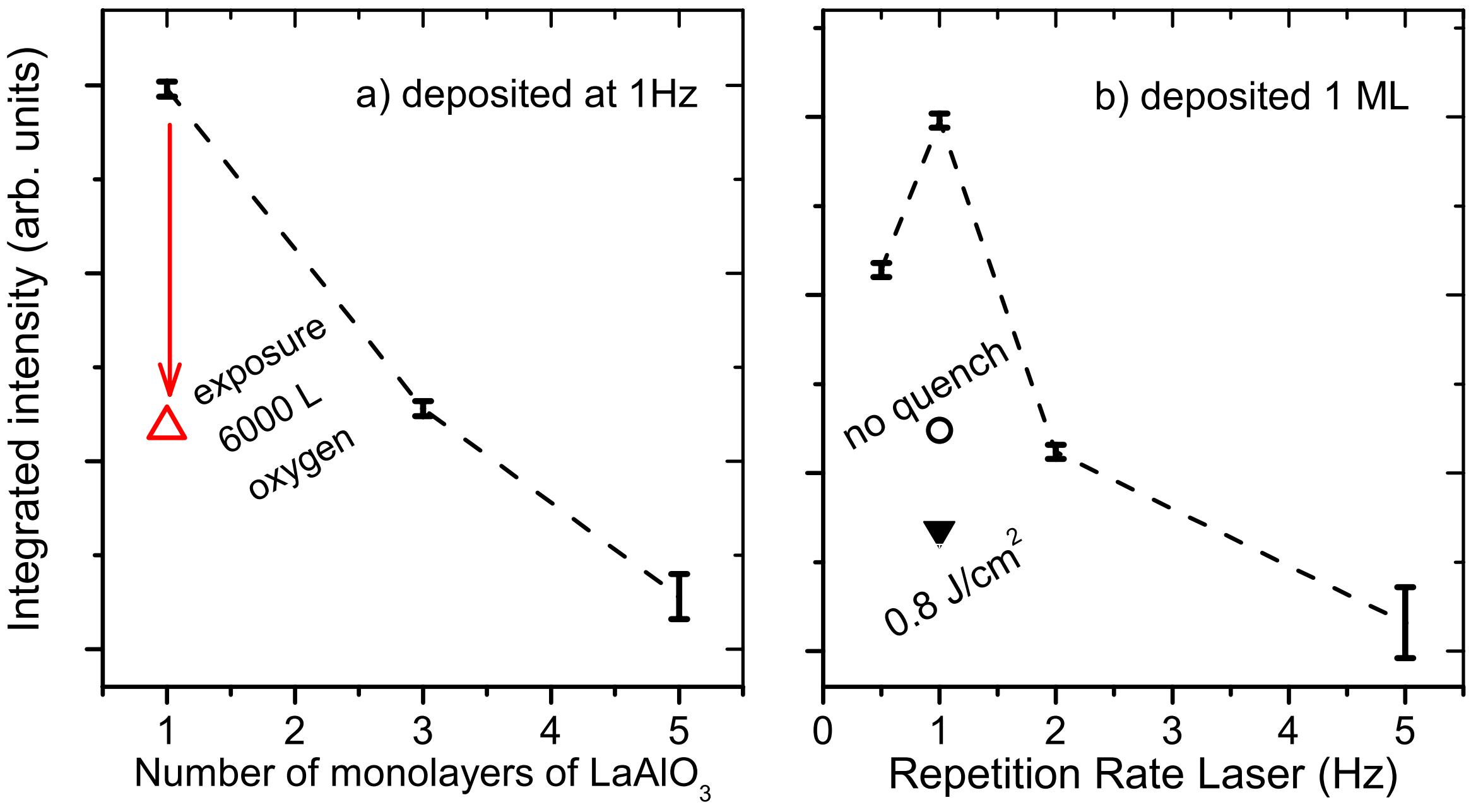

Figure 6. (color online) Analysis of UPS data for different samples. a) Comparison of the UPS signal near $E_F$ for different $LaAlO_3$ film thicknesses by integration of the difference intensity (for example from Fig. 5); the error bars indicate the statistical error; the connecting line is a guide to the eye; the red open triangle is the value found for the 1 Hz, 1 ML sample after exposure to ~ 6000L of oxygen. b) Comparison of the integrated difference intensity as a function of the laser repetition rate (1 ML --- 1 Hz, 2 Hz, 5 Hz), where the error bars indicate the statistical error; the connecting line is a guide to the eye; the open circle is the value found for a sample where the pressure quench described in the text is not applied; the solid triangle represents the value found for a sample made when a laser energy of 0.8 J $cm^{-2}$ per pulse is used.